\documentclass[10pt,twoside]{article}

\usepackage{fancyhdr}
\usepackage[T1]{fontenc}
\usepackage[utf8]{inputenc}

\usepackage{txfonts}
\usepackage[svgnames]{xcolor}
\usepackage[paperheight=27.94cm,paperwidth=21.59cm,left=2.54cm,right=2.54cm,top=2.54cm,bottom=2.54cm]{geometry}

\usepackage{titlesec}
\titleformat*{\section}{\normalsize\bfseries}
\titleformat*{\subsection}{\normalsize\bfseries\itshape}
\titleformat*{\subsubsection}{\normalsize\itshape}

\usepackage{graphicx}
\usepackage{placeins}
\usepackage{framed}
\usepackage{float}
\usepackage{subcaption}
\usepackage{adjustbox}
\usepackage{multicol}
\usepackage{multirow}

\usepackage{url}
\usepackage{hyperref}
\usepackage{doi}
\hypersetup{
    colorlinks = true,
    urlcolor   = blue,
    citecolor  = blue, 
    linkcolor = blue
}
\usepackage[capitalise]{cleveref} 

\usepackage[bibstyle=ieee, date=year, citestyle=numeric-comp, url=true, natbib=true, mincitenames=1, maxcitenames=2, uniquelist=false, uniquename=false]{biblatex}
\AtEveryBibitem{%
    \clearfield{extra}
    \clearfield{urlyear}
    \clearfield{urlmonth}
    \clearfield{note}%
    \clearlist{language}%
    \clearfield{issn}%
    \clearfield{isbn}%
}

\usepackage[hang, flushmargin, symbol]{footmisc}
\usepackage{enumitem}

\title{Experimental evaluation of advanced control strategies for high-blockage cross-flow turbine arrays}
\author{Overleaf}
\date{\today}

\begin{document}\thispagestyle{empty} 

\begin{center}
    \begin{figure}[h]
        \centering
        \includegraphics[width = 0.2\textwidth]{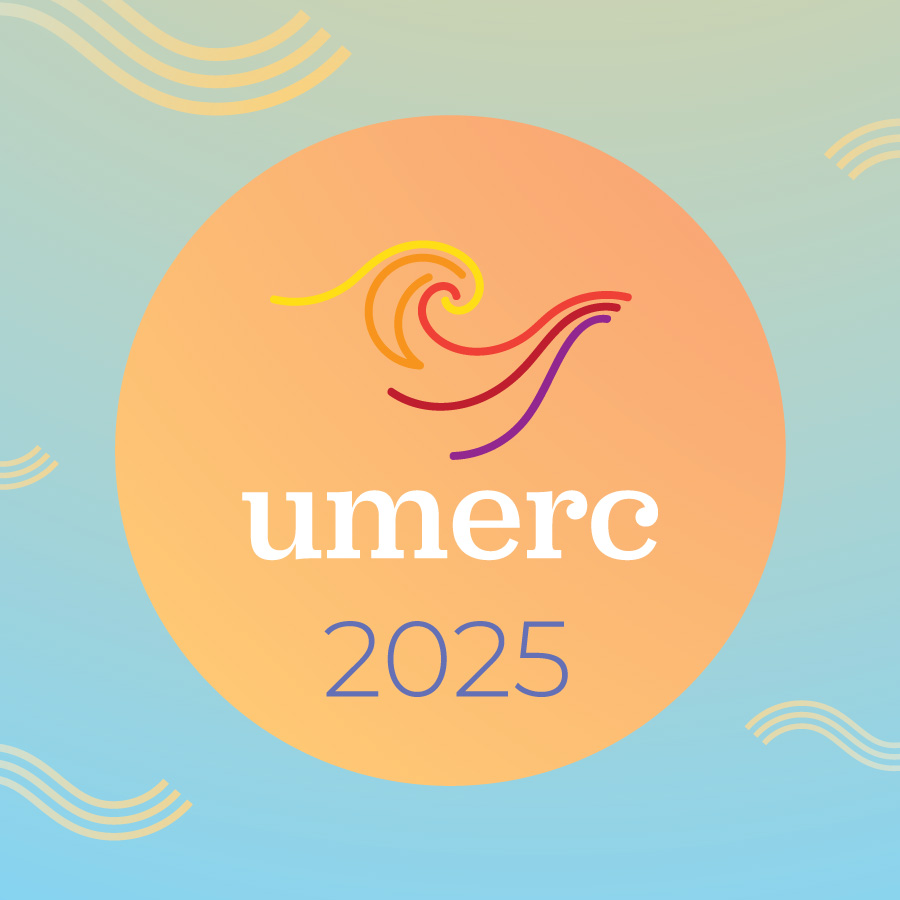} \\
    \end{figure}
    \huge
    UMERC+OREC \\
    2025 Conference \\
    \vspace{1\baselineskip}
    \large
    \textit{12-14 August | Corvallis, OR USA} \\
    \vspace{1\baselineskip}
       
    \huge
    Experimental evaluation of advanced control strategies for high-blockage cross-flow turbine arrays
    \vspace{1\baselineskip}
    
    \Large
    Aidan Hunt\footnotemark[1], Gregory Talpey, and Brian Polagye 

    \footnotetext[1]{Corresponding author. Email address: ahunt94@uw.edu}
    \footnotetext[0]{This work was supported by the United States Advanced Research Projects Agency – Energy (ARPA-E) under award number DE-AR0001441.}
    \normalsize
    \begin{center}
        \textit{Department of Mechanical Engineering, University of Washington, Seattle, WA, USA}
    \end{center}
\end{center}         

\noindent\rule{\textwidth}{0.4pt}
\section*{Abstract}
In river or tidal channels, cross-flow turbines can achieve higher blockage ratios than other turbine variants, and are therefore able to achieve higher efficiencies.
Here, we experimentally investigate how array control strategies might further influence the efficiency of a high-blockage dual-rotor cross-flow turbine array.
Array performance is evaluated under coordinated constant speed control, uncoordinated torque control, and coordinated intracycle speed control at blockage ratios of $35\% - 55\%$. 
In contrast to prior work at lower blockage, the evaluated control strategies do not yield significant improvements in efficiency and intracycle control is found to generally reduce array performance.
While these results suggest limited benefits to more advanced control strategies at high blockage, this has the benefit of simplifying the system design space for array-level control.

\vspace{1\baselineskip}
\textit{Keywords:} cross-flow turbine; vertical-axis turbine; array control; blockage; intracycle control

\noindent\rule{\textwidth}{0.4pt}

\setlength{\parindent}{.17in} 

\section{Introduction}

The performance of a row of turbines operating in a river or tidal channel is influenced by the fraction of the channel the turbines occupy.
In these channels, resistance to flow through the turbines, combined with confinement from the channel boundaries, accelerates the flow through the rotor, which increases each turbine's efficiency \citep{garrett_efficiency_2007, nishino_efficiency_2012, houlsby_power_2017}.
While confinement augments the performance of all turbine designs, cross-flow turbines are particularly well-suited to exploiting this effect, since their rectangular projected area allows them to be densely arranged in river or tidal channels, which have a fundamentally rectangular form factor.
The portion of the channel occupied by a row of turbines is quantified by the ``blockage ratio'', which is the ratio between the total turbine swept area and the channel cross-sectional area.
For a row of cross-flow turbines, this is given as
\begin{equation}
    \beta = \frac{nHD}{wh} \ ,
\end{equation}
where $H$ is the blade span, $D$ is the turbine diameter, $n$ is the number of turbines, and $w$ and $h$ are the channel width and depth, respectively.

Within a row of cross-flow turbines, the kinematics of adjacent rotors can be coordinated with one another to influence performance.
At lower blockage ratios, both rotation scheme (i.e., whether the rotors rotate in the same direction or opposite direction) and the phase offset between the rotational cycles of adjacent turbines have been shown to influence array efficiency and wake characteristics \citep{strom_advanced_2018, posa_wake_2019, scherl_geometric_2020, jin_aerodynamic_2020, gauvin-tremblay_hydrokinetic_2022, scherl_wake_2025}.
Additionally, for individual rotors, intracycle control strategies—wherein the angular velocity of a turbine oscillates as a function of phase throughout a rotational cycle—have also been shown to enhance efficiency at lower blockage ratios \citep{strom_intracycle_2017, dave_simulations_2021, athair_intracycle_2023}.
However, none of these strategies have been evaluated at the upper end of blockage ratios achievable in a realistic channel (i.e., $30\% \leq \beta \leq 60\%$).
In this work, three array control strategies are applied to a laboratory-scale dual-rotor cross-flow turbine array: coordinated constant speed control, uncoordinated torque control, and coordinated intracycle speed control.
Array performance is evaluated for each control strategy over a range of tip-speed ratios at $\beta = 35\%-55\%$.

\section{Methods}

The test facility and experimental setup are identical to that employed in related studies by the authors of high-blockage cross-flow turbine arrays \citep{hunt_experimental_2023, hunt_experimental_2024a, hunt_performance_2025}, to which the reader is referred for detailed descriptions and schematics.
The flow conditions corresponding to $\beta = 35\%$, $45\%$, and $55\%$ are the same as described in \citep{hunt_performance_2025} and the geometry of each two-bladed rotor is the same as described in \citep{hunt_experimental_2024a}.

\subsection{Control Strategies}

\begin{figure*}
    \begin{minipage}{0.525\textwidth}
        \centering
        \includegraphics[width=\textwidth]{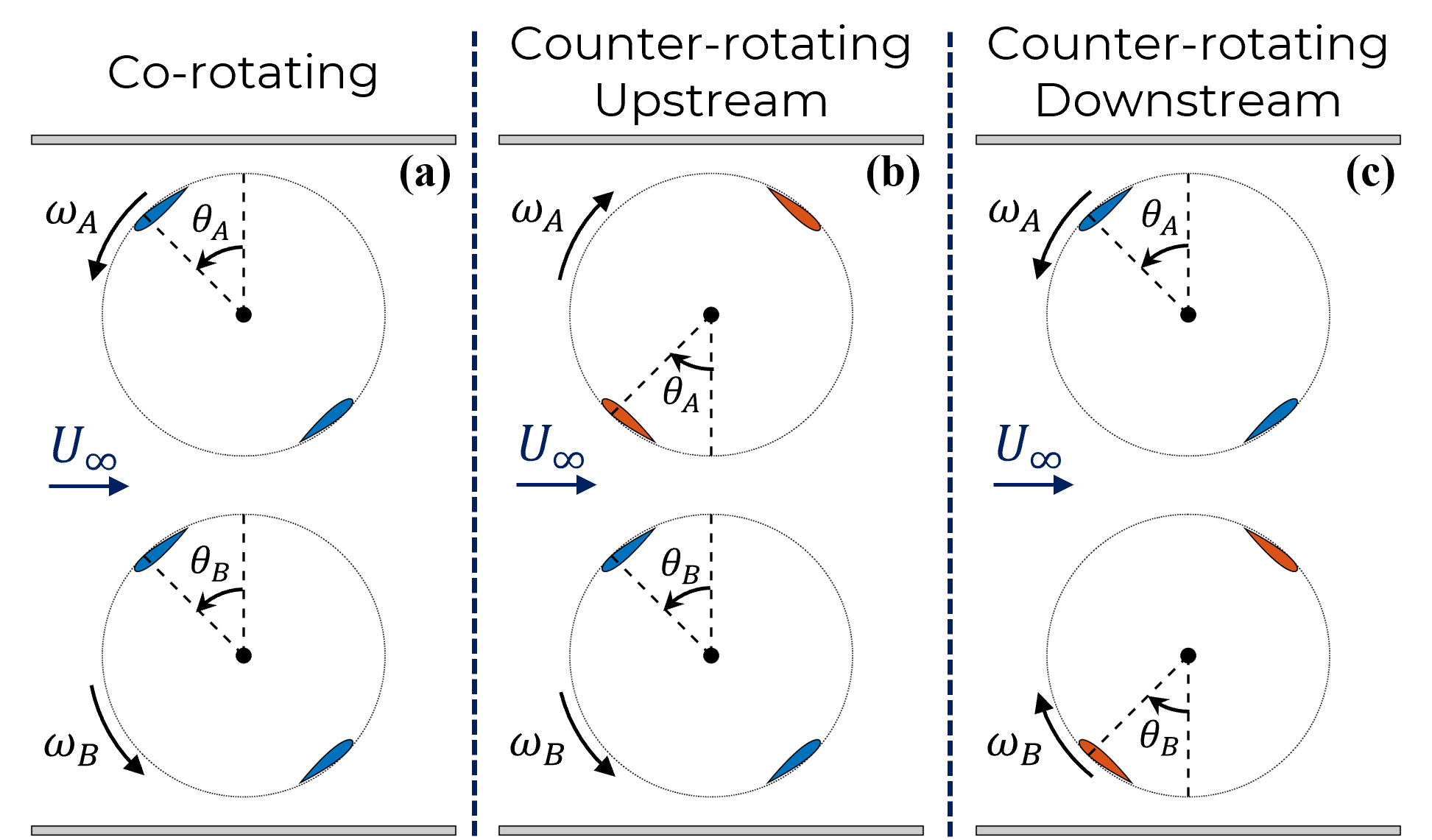}
         {\phantomsubcaption\label{fig:corotate}
          \phantomsubcaption\label{fig:counterup}
          \phantomsubcaption\label{fig:counterdown}}
        \caption{Different rotation schemes for a two-rotor cross-flow turbine array.}
        \label{fig:rotationSchemes}
    \end{minipage} 
    \hfill
    \centering
    \begin{minipage}{0.425\textwidth}
        \centering
        \includegraphics[width=\textwidth]{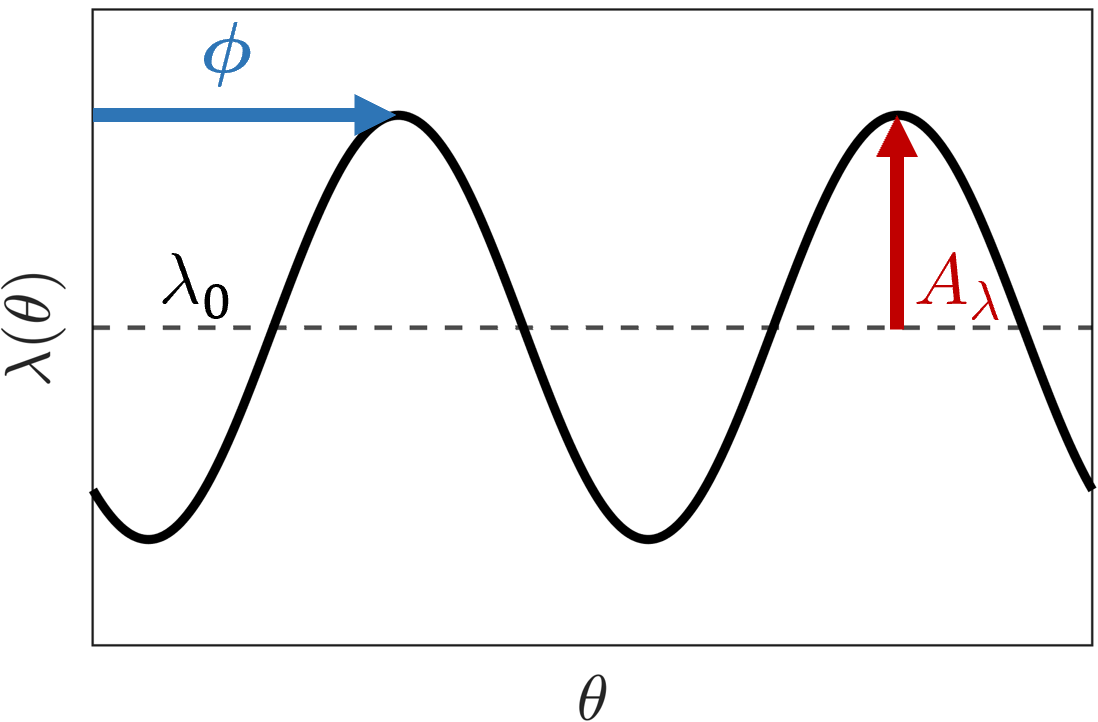}
        \caption{Example of a sinusoidal intracycle velocity trajectory for a two-bladed turbine resulting from \Cref{eq:intracycle}.}
        \label{fig:intracycleSchematic}
    \end{minipage} 
\end{figure*}

For a dual-turbine array such as that considered in this study, the turbines can either rotate in the same direction (``co-rotating''; \cref{fig:corotate}) or in opposite directions (``counter-rotating'').
Counter-rotating schemes can be further distinguished as either ``counter-rotating upstream'' (\cref{fig:counterup}) or ``counter-rotating downstream'' (\cref{fig:counterdown}), depending on whether adjacent blades are translating upstream or translating downstream when passing through the array centerline.
Within a given rotation scheme, the turbines in the array can either be operated independently of one another (i.e., ``uncoordinated control'') or coordinated such that a specified relationship between the angular velocities, torques, or the phase difference between the turbines the turbines is maintained.
Here, we evaluate the array’s performance under several control strategies for these rotation schemes, with an emphasis on counter-rotating downstream arrays (\cref{fig:counterdown}) since prior work at lower blockage \citep{gauvin-tremblay_hydrokinetic_2022, scherl_wake_2025} has suggested that this rotation scheme yields the highest array performance.
For all control schemes, the rotation rate of each turbine is non-dimensionalized as the tip-speed ratio,
\begin{equation}
    \lambda = \frac{\omega R}{U_{\infty}}
\end{equation}
where $\omega$ is the angular velocity of the turbine, $R$ is the turbine radius, and $U_{\infty}$ is the freestream velocity.

\textit{Coordinated Constant Speed Control}:
Under coordinated constant speed control, both turbines in the array rotate at the same, constant angular velocity, which is prescribed by their respective servomotors.
The kinematics of the two turbines are further coordinated with one another by maintaining a constant phase difference between their rotational cycles, defined as $\Delta \theta = \theta_A - \theta_B$, where $\theta_A$ and $\theta_B$ are the azimuthal positions of the turbine blades as shown in \cref{fig:rotationSchemes}.
Phase differences of $\Delta \theta = 0^{\circ}$, $\pm45^{\circ}$, and $90^{\circ}$, are considered in this study.
These phase differences span the full range of unique phase differences for a two-bladed turbine, and are expected to be representative of the array's sensitivity to $\Delta \theta$ based on the results of \citet{scherl_wake_2025}. 
The target phase difference is maintained within $\pm 0.1^{\circ}$ by a closed-loop controller.
Array performance under this control strategy was characterized for co-rotating, counter-rotating upstream, and counter-rotating downstream schemes over a range of $\lambda$ at each $\beta$.

\textit{Uncoordinated Constant Torque Control}:
Under constant torque control, a constant resistive torque (i.e., torque opposing rotation) is applied to each turbine by its servomotor.
Here, we consider uncoordinated constant torque control, meaning that both the angular velocities of the turbines and the phase difference between them are not explicitly prescribed.
Because of this, the phase difference would be expected to randomly progress in time as a consequence of turbulence.
At each $\beta$, the performance of a counter-rotating downstream array was characterized over a range of $\lambda$ by varying the applied torque, with the array tip-speed ratio taken as the average of the turbine tip-speed ratios. 

\textit{Coordinated Intracycle Speed Control}:
Under intracycle speed control, the prescribed angular velocity of each turbine varies throughout a rotational cycle.
Here, we consider a sinusoidal velocity trajectory as a function of blade position, formulated by \citet{athair_intracycle_2023} as
\begin{equation}
    \lambda(\theta) = \lambda_0 + A_{\lambda}\cos(N(\theta - \phi))
    \label{eq:intracycle}
\end{equation}
\noindent where $\lambda_0$ is the phase-average tip-speed ratio, $A_{\lambda}$ is the oscillation amplitude of $\lambda$ within a rotational cycle, $N$ is the number of blades, and $\phi$ is the azimuthal position corresponding the maximum angular velocity.
The influences of $\lambda_0$, $A_{\lambda}$, and $\phi$ on the resulting velocity trajectory are qualitatively illustrated in \cref{fig:intracycleSchematic}.
However, we note that the \textit{time-average} $\lambda$ decreases with $A_\lambda$ since, within a given rotational cycle, the turbine spends more time moving at lower rotation rates \citep{athair_intracycle_2023}.
Here, the performance of a counter-rotating downstream array is characterized at each $\beta$ for several combinations of $A_\lambda$ ($[0.1\lambda_0, \  0.8\lambda_0]$ in increments of $0.1\lambda_0$), and $\phi$ ($[0^{\circ}, 180^{\circ})$ in increments of $22.5^{\circ}$) at the $\lambda_0$ corresponding to maximum efficiency under constant speed control.
Additionally, the turbines in the array are coordinated with one another to maintain a constant phase difference of $\Delta \theta = 0^{\circ}$.

\subsection{Performance Metrics}

For each control strategy, the time-average mechanical efficiency and streamwise loading of the array are quantified via the coefficients of performance and thrust, calculated respectively as

\noindent\begin{minipage}{0.5\textwidth}
    \begin{equation}
        \langle C_{P} \rangle = 
        \frac{ \langle \, Q_{H,A}\omega_{A} + Q_{H,B}\omega_{B} \, \rangle} {\frac{1}{2}\rho \,  \langle U_{\infty}^3 \rangle \,  2HD} \ \ ,
        \label{eq:cp}
    \end{equation}
\end{minipage}%
\begin{minipage}{0.5\textwidth}
    \begin{equation}
    \langle C_{T} \rangle = \frac{ \langle \, T_A + T_B \, \rangle }{\frac{1}{2} \rho \, \langle  U_{\infty}^2 \rangle \,  2HD} \ \ ,
    \label{eq:ct}
    \end{equation}
\end{minipage}
where $\omega_{A}$ and $\omega_{B}$ are the angular velocities of each turbine, $Q_{H,A}$ and $Q_{H,B}$ are the hydrodynamic torques on each turbine, $T_A$ and $T_B$ are the streamwise forces on each turbine, $\rho$ is the density of the freestream flow, and $\langle \ \rangle$ denotes averaging in time.
We note that measurements of $U_{\infty}^2(t)$ and $U_{\infty}^3(t)$ are not synchronized with measurements of torque and rotation rate, and are therefore averaged separately as indicated in \cref{eq:cp,eq:ct}.
The hydrodynamic torque on each turbine ($Q_H$) is related to the net control torque applied by the servomotor and bottom bearing ($Q_C$) by Newton's second law for rotation,
\begin{equation}
    Q_H = Q_C - J\dot{\omega} \ \ ,
\end{equation}
where $\dot{\omega}$ is the angular acceleration and $J$ is the moment of inertia.
Under constant speed control, $\dot{\omega} \approx 0$ such that $Q_H = Q_C$.
However, under constant torque control and intracycle speed control, $\omega$ varies in time, such that the $Q_H$ and $Q_C$ are not necessarily equal.
Therefore, for these rotation schemes $Q_H$ is calculated by subtracting $J\dot{\omega}$ from the measured $Q_C$, with $J$ estimated by rotating the turbines in air as described in Refs. \citep{strom_intracycle_2017, polagye_comparison_2019, athair_intracycle_2023}.


\section{Results}


\begin{figure}[t]
    \centering
    \includegraphics[width=\textwidth]{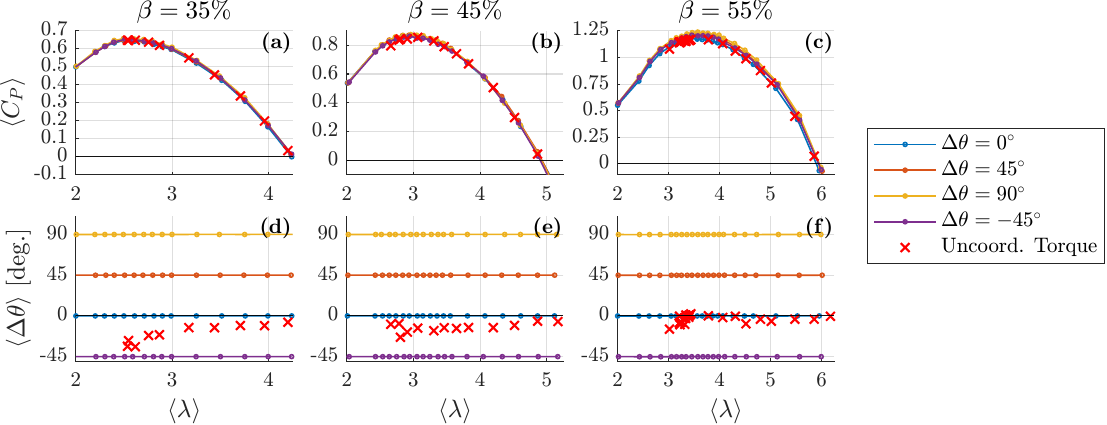}
    {\phantomsubcaption\label{fig:cp_ds_35}
    \phantomsubcaption\label{fig:cp_ds_45}
    \phantomsubcaption\label{fig:cp_ds_55}
    \phantomsubcaption\label{fig:phasediff_ds_35}
    \phantomsubcaption\label{fig:phasediff_ds_45}
    \phantomsubcaption\label{fig:phasediff_ds_55}
    }
    \caption{\subref{fig:cp_ds_35} - \subref{fig:cp_ds_55}: counter-rotating downstream array efficiency as a function of tip-speed ratio and phase difference for $\beta = 35\%$, $45\%$, and $55\%$. Note the different axes limits for each $\beta$. \subref{fig:phasediff_ds_35} - \subref{fig:phasediff_ds_55} measured phase difference for each case. The measured $\langle C_P \rangle$ and $\langle \Delta \theta \rangle$ of the array under uncoordinated torque control are denoted as red `x' symbols.}
    \label{fig:cp_thetaDiff_downstream}
\end{figure}

\begin{figure}[h]
    \centering
    \includegraphics[width=\textwidth]{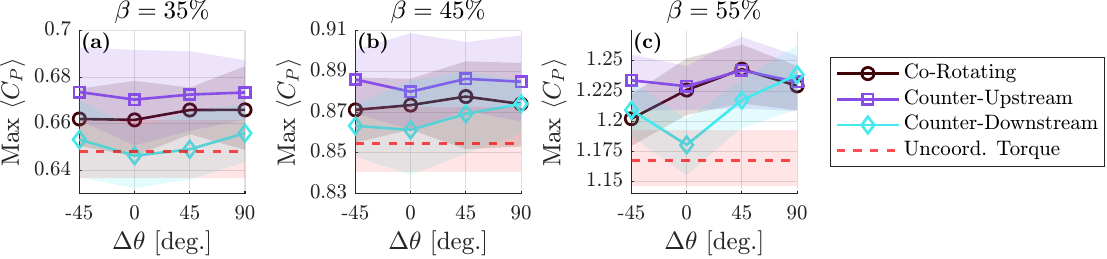}
    \caption{Maximum time-average $C_P$ for each rotation scheme under coordinated constant speed control. The dashed red line indicates the maximum $\langle C_P \rangle$ observed for a counter-rotating downstream array under uncoordinated constant torque control. The shaded regions indicate the interquartile range of cycle-average $C_P$ at the optimal operating point.}
    \label{fig:cpMax_thetaDiff}
\end{figure}

The time-average efficiency of the array under coordinated constant speed control is shown for the counter-rotating downstream case as a function of $\lambda$ in \cref{fig:cp_thetaDiff_downstream} for each combination of $\beta$ and $\Delta \theta$.
At each tested $\beta$, the highest $\langle C_P \rangle$ is achieved with $\Delta \theta = 90^{\circ}$, which is in agreement with the observations of \citet{scherl_wake_2025} at $\beta = 20\%$.
However, the influence of $\Delta \theta$ on performance is subtle relative to that of blockage and the tip-speed ratio.
Although the influence of $\Delta\theta$ appears to increase slightly with $\beta$, the maximum $\langle C_P \rangle$ observed varies across the tested $\Delta \theta$ by $<2\%$ at $\beta = 35\%$ and $45\%$, and $<5\%$ at $\beta = 55\%$.

Similar trends in maximum $\langle C_P \rangle$ with phase difference are observed for both the counter-rotating upstream and co-rotating arrays, as well (\cref{fig:cpMax_thetaDiff}).
Of the three rotation schemes evaluated, the counter-rotating downstream array exhibits the most dependence on $\Delta \theta$, which is consistent with observations by \citet{scherl_wake_2025}.
However, in contrast to the results of \citet{scherl_wake_2025}, at all tested $\beta$ the highest absolute performance at is obtained with a counter-rotating \textit{upstream} scheme, rather than a counter-rotating \textit{downstream} scheme.
Nonetheless, at a given $\beta$ the percent difference in maximum $\langle C_P \rangle$ across all combinations of rotation scheme and $\Delta \theta$ evaluated is, at most, $\sim \! 5\%$, and there is significant overlap in the interquartile ranges of the cycle-average $C_P$ observed for each scheme (\cref{fig:cpMax_thetaDiff}).
Consequently, the choice of coordinated constant speed control scheme does not significantly affect array performance.
This result is actually beneficial for confined arrays, given that a continuous fence would necessarily alternate counter-rotating directions.


For the counter-rotating downstream array, operation under uncoordinated torque control yields similar time-average behavior in $C_P$ at the tested $\beta$ as for coordinated constant speed control  (\cref{fig:cp_ds_35,fig:cp_ds_45,fig:cp_ds_55}).
Notably, the measured phase difference under uncoordinated torque control (\cref{fig:phasediff_ds_35,fig:phasediff_ds_45,fig:phasediff_ds_55}) tends toward $\Delta \theta = 0^{\circ}$  as $\lambda$ increases, with this trend becoming stronger at higher $\beta$.
In other words, at the tested $\beta$, the turbines in the array self-synchronize to $\Delta \theta = 0^{\circ}$ despite operating under uncoordinated constant torque control.
This observation is consistent with the maximum $ \langle C_P \rangle $ obtained for uncoordinated torque control, which is generally comparable or slightly less than that under constant speed coordinated control with $\Delta \theta = 0^{\circ}$ (\cref{fig:cpMax_thetaDiff}).
We hypothesize that the somewhat lower $\langle C_P \rangle$ of uncoordinated torque control is a result of unsteady variations in $\lambda$, due in part to fluctuations in the freestream flow.


\begin{figure}[t]
    \centering
    \includegraphics[width=\textwidth]{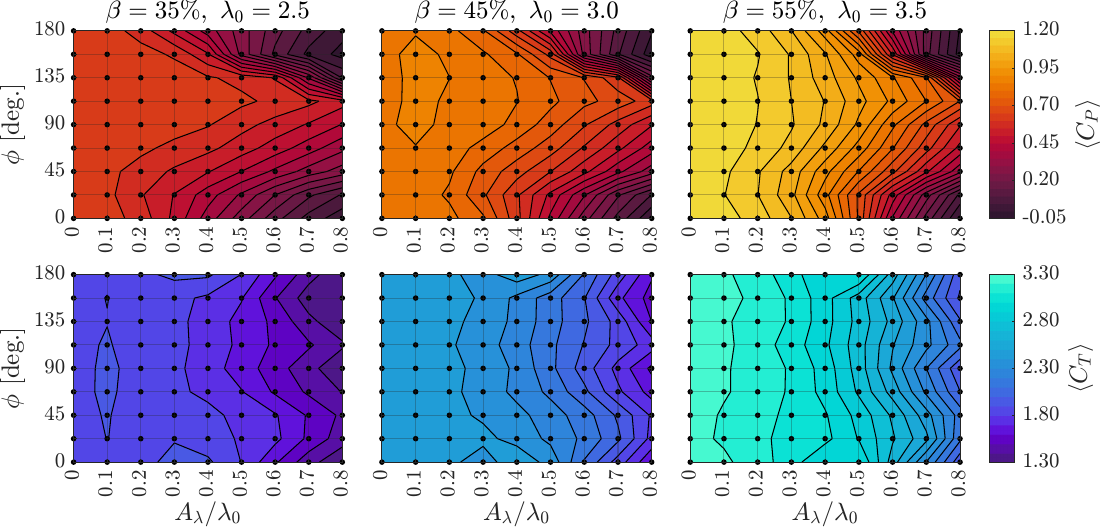}
    \caption{Time-average $C_P$ (top row) and $C_T$ (bottom row) for a counter-rotating downstream array operating under various coordinated intracycle speed control trajectories. Each grid point represents a particular combination of $A_\lambda$ and $\phi$ that was tested at a given $\beta$ and $\lambda_0$. The constant speed control reference case is given by $A_\lambda / \lambda_0 = 0$.}
    \label{fig:intracycle_results}
\end{figure}

\Cref{fig:intracycle_results} shows the contours of the time-average $C_P$ and $C_T$ for the counter-rotating downstream array operating under various coordinated intracycle speed control trajectories.
In general, coordinated intracycle control provides little-to-no benefit over constant speed control for this range of $\beta$.
At a given $\lambda_0$, time-average $C_P$ tends to decrease with increasing oscillation amplitude, with excursions from $\phi \approx 120^{\circ}$ resulting in further reductions to $ \langle C_P \rangle$.
While a slight increase in $\langle C_P \rangle $ is observed at $\beta = 45\%$ and $\lambda_0 = 3.0$ for a range of $\phi$ at $A_\lambda / \lambda_0 = 0.1$, the benefit is minor.

We hypothesize that reduced performance under intracycle control is related to reductions in $\langle C_T \rangle$.
Specifically, as $A_\lambda / \lambda_0$ increases, even though the maximum instantaneous $\lambda$ increases, the time-average $\lambda$ decreases \citep{athair_intracycle_2023}.
Consequently, under intracycle control the array is, on average, more porous to the flow than under constant speed control, leading to a reduction in thrust.
Since performance in confined flow increases with both $\beta$ and $C_T$ \citep{garrett_efficiency_2007, houlsby_power_2017}, the reduction in thrust from intracycle control undermines the mechanism for blockage-driven performance augmentation.
While not shown, we note that technoeconomic modeling indicates that a corresponding reduction in the power-to-force ratio via intracycle control does not yield a meaningful reduction in levelized cost of energy.


\section{Discussion and Conclusions}
Overall, this study identified limited benefits to more advanced control strategies for cross-flow turbines in high-blockage flows.
For the array considered in this study, performance is primarily influenced by the blockage and the tip-speed ratios, with coordination between the turbines having only a minor effect.
Additionally, the oscillating kinematics imposed by intracycle control appear to undercut beneficial interactions between the turbine and the channel.
From a systems design perspective, the additional complexity necessary to implement these control strategies would have an economic cost disproportionate to the limited benefits.
In contrast, optimizing the rotor geometry for high confinement is a simpler and more effective means of exploiting high-blockage dynamics \citep{kong_correction_2025, hunt_performance_2025}.
Nonetheless, our results indicate that high-blockage cross-flow turbine arrays perform sufficiently well under simple control schemes, which provides a welcome reduction of the system design space.

\FloatBarrier
\newpage
\printbibliography

\end{document}